\documentclass[aps,prl,reprint, superscriptaddress,nofootinbib]{revtex4-2}
\usepackage{graphicx}
\usepackage{dcolumn}
\usepackage{color}
\usepackage{bm}
\usepackage{amsmath, siunitx}
\usepackage{ulem}
\usepackage{gensymb}
\usepackage{float}
\usepackage{siunitx}

\newcommand{\nitase}{Ni$_{1/4}$TaSe$_2$ }

\newcommand{\ku}{K$_{\text{u}}$}
\newcommand{\bw}{B$_{\text{w}}$}
\newcommand{\rhoxy}{$\rho_{\text{xy}}$ }
\newcommand{\rhoxx}{$\rho_{\text{xx}}$ }

\begin{document}

\title{Electrically controlled Heat Assisted Magnetic Recording in Intercalated 2D Magnets}

\author{Josue Rodriguez}
\affiliation{Department of Physics, University of California, Berkeley, California 94720, USA}
\affiliation{Materials Science Division, Lawrence Berkeley National Laboratory, Berkeley, California 94720, USA}
\author{Ruishi Qi}
\affiliation{Department of Physics, University of California, Berkeley, California 94720, USA}
\affiliation{Materials Science Division, Lawrence Berkeley National Laboratory, Berkeley, California 94720, USA}
\affiliation{Kavli Energy NanoScience Institute at the University of California, Berkeley
and the Lawrence Berkeley National Laboratory, Berkeley, California 94720, USA}

\author{Catherine Xu}
\affiliation{Department of Physics, University of California, Berkeley, California 94720, USA}
\affiliation{Materials Science Division, Lawrence Berkeley National Laboratory, Berkeley, California 94720, USA}

\author{Feng Wang}
\affiliation{Department of Physics, University of California, Berkeley, California 94720, USA}
\affiliation{Materials Science Division, Lawrence Berkeley National Laboratory, Berkeley, California 94720, USA}
\affiliation{Kavli Energy NanoScience Institute at the University of California, Berkeley
and the Lawrence Berkeley National Laboratory, Berkeley, California 94720, USA}
\author{James G. Analytis}
\altaffiliation{Contact for correspondence, analytis@berkeley.edu}
\affiliation{Department of Physics, University of California, Berkeley, California 94720, USA}
\affiliation{Materials Science Division, Lawrence Berkeley National Laboratory, Berkeley, California 94720, USA}
\affiliation{Kavli Energy NanoScience Institute at the University of California, Berkeley
and the Lawrence Berkeley National Laboratory, Berkeley, California 94720, USA}
\author{Hossein Taghinejad}
\altaffiliation{Contact for correspondence, h.taghinejad@berkeley.edu}
\affiliation{Department of Physics, University of California, Berkeley, California 94720, USA}
\affiliation{Kavli Energy NanoScience Institute at the University of California, Berkeley, California 94720, USA}


\begin{abstract}
The ever-increasing demand for fast, reliable, and energy-efficient information storage continues to push magnetic memory technologies toward their fundamental limits. Conventional scaling strategies, which rely on reducing bit size, inevitably run into the “magnetic recording trilemma,” where signal-to-noise ratio, thermal stability, and writability cannot all be optimized simultaneously. Heat-assisted magnetic recording (HAMR) has emerged as the leading solution, enabling high-density storage by transiently heating the medium during the write cycle. However, the reliance on laser optics and plasmonic transducers restricts HAMR primarily to hard-disk drives, limiting its integration with on-chip or embedded architectures. Here, we demonstrate an electronic variant of HAMR in which Joule heating from low-current density current pulses facilitates data writing, while the anomalous Hall effect provides electronic readout. Employing intercalated 2D magnet \nitase, we show direct evidence that current pulses heat the material above its Curie temperature, during which a small magnetic field of $\sim$ \SI{2}{\milli\tesla} (100 times smaller than the coercive field) enables efficient data writing. The all-electronic approach combined with the 2D magnetic medium creates timely opportunities to revisit the energy-assisted magnetization recording, enabling new recording schemes that combine fundamental novelty with technological impact.
\end{abstract}

\maketitle

\section{Introduction}
The exponential growth of digital data underpins nearly every aspect of modern life, from cloud computing and artificial intelligence to scientific discovery and personal electronics. Meeting this demand requires storage technologies that are not only scalable in density but also energy-efficient and reliable over long timescales. Magnetic and spintronic memories stand out in this regard as the intrinsic non-volatility of magnetization offers low-energy data retention, \cite{Puebla2020SpintronicDevices,Nguyen2024_SOTMRAM_Progress, 10.1063/1.3567780, Duffee2025}  while their compatibility with fast electronic readout makes them attractive for high-performance applications. Yet, realizing their full potential depends critically on discovering and engineering new materials that can balance thermal stability, writability, and integration with emerging device architectures. In this context, quantum materials and particularly emerging two-dimensional (2D) magnets  with tunable anisotropy and Curie temperatures provide unique opportunities to be used as the recording media. The structural versatility and tunable magnetic interactions in 2D magnets expand the design space well beyond conventional alloys, enabling exploration of recording schemes that combine fundamental novelty with technological advancements.

\begin{figure}[t]
\includegraphics[width=0.9\columnwidth]{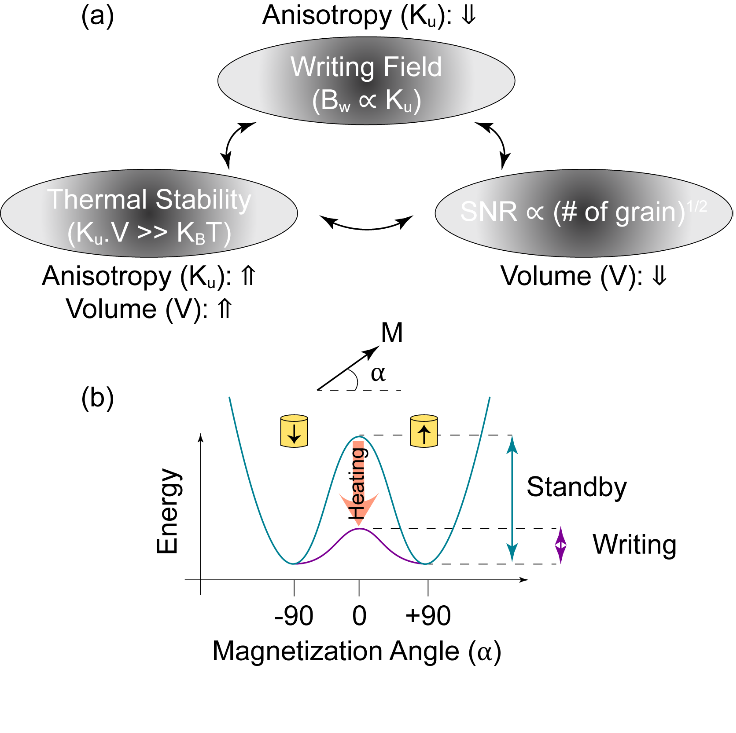}
    \caption{\textbf{Magnetic recording trilemma and HAMR concept.}
    \textbf{a} The trilemma of magnetic recording: increasing areal density by shrinking bit size imposes contradictory requirements for simultaneously maintaining reasonable SNR (depends on volume, V), thermal stability (depends on anisotropy, \ku, and V), and writing field (\bw, depends on \ku). Optimizing one metric necessarily compromises another. 
    \textbf{b} HAMR circumvents this limitation: transiently heating the sample above its Curie temperature reduces the anisotropy barrier, enabling bit flipping with modest fields during the write cycle. Upon cooling, the magnetic anisotropy is restored to a large value, ensuring thermal stability for long-term data retention. In our work, Joule heating from short current pulses serves as the source of transient temperature rise.
    }
    \label{fig:trilema}
\end{figure}

At the center of every magnetic storage technology lies its areal density, which determines how much information can be encoded within a given surface area and ultimately sets the limit for storage capacity. To increase the areal density of the recording media, we need to reduce the size of magnetic bits. This seemingly obvious strategy, however, faces a trilemma that we schematically illustrate in Figure \ref{fig:trilema}(a). The trilemma stems from simultaneously satisfying three major performance measures: signal-to-noise-ratio (SNR), thermal stability, and the magnetic field used for writing (\bw, a measure of writability). In reality, each data bit is a granular area composed of multiple grains, making the boundary between adjacent bits rather rough (not a straight line). As such, to maintain high SNR while shrinking the bit size, the number of grains per bit should be kept constant ($\sim$10), or the edge roughness reduces the SNR \cite{Plumer2011_NewParadigms,Meo2023_SNR_HAMR}. Thus, reducing the bit size requires reducing the grain volume (V). However, reducing the volume increases the risk of data corruption as the magnetization becomes susceptible to thermal instability (i.e., random bit flipping) due to the superparamagnetic effects \cite{White1999_SuperparamagneticLimit}. To avoid this issue, the magnetic anisotropy (\ku) of grains should be increased so that the energy barrier for bit flipping (proportional to K$_{\text{u}}\times$V) is sufficiently above the thermal energy (k$_{\text{B}}$T) \cite{Weller1999_ThermalLimits}. This leads to third tension: increasing anisotropy directly increases the writing field (\bw) required for the magnetization reversal because the coercivity of the grains scale with the magnitude of the anisotropy, \bw $\propto$ \ku. \cite{Kryder2008_HAMR_Trilemma, doi:10.1021/acs.nanolett.4c06592} As such, satisfying all three performance measures, while shrinking the bit size, becomes virtually impossible, hence the trilemma.

To break out of the trilemma, the concept of energy-assisted magnetic recording (EAMR) is introduced and different variants of it have been developed \cite{Miao2015_EAMR_OpticalDelivery,Weller2016_FePt_HAMR,Toshiba2017_MAMR}. In EAMR, the idea is to maintain high SNR and robust thermal stability via designing magnetic materials with large K$_{u}$ and small V, but to alleviate the large \bw\,penalty, an external source of energy is supplied during the data-writing cycle to assist with the magnetization reversal (Figure \ref{fig:trilema}(b)), hence satisfying the third performance measure.  The extra energy is often provided in the form of either microwave \cite{Toshiba2017_MAMR} or heat \cite{Weller2016_FePt_HAMR}, the latter of which has gained widespread adoption and lead to the heat-assisted magnetization recording (HAMR). \cite {7115916} In the HAMR approach, a laser pulse is used to heat the magnetic media to above its Curie temperature, which temporarily reduces the magnetic anisotropy, hence allowing writing (i.e., reversing) the magnetization direction using small writing fields, \bw. Following the removal of the laser pulse, the medium restores its large anisotropy for stable data retention. To go beyond the diffraction-limitation of the light, the laser is focused on a plasmonic transducer which enables confining the heat to a smaller spot size, roughly the lateral size of a single bit of data \cite{Kief2018_HAMR_Materials}. Although the read-head of hard disk drive can accommodate such bulky additions (lasers, plasmonic transducers, etc.), a similar setup is not appealing for integrated solutions and embedded memory architectures.

Here, we demonstrate an electronic variant of HAMR, in which magnetization reversal is assisted not by laser heating but by Joule heating from current pulses with densities as low as $10\times10^{4}$\si{\ampere}.\si{\per\centi\meter\squared}. Combined with electronic readout via the anomalous Hall effect, our approach enables an all-electronic read-write scheme for magnetization recording. We combine electronic transport measurements and magneto-optical imaging and provide unambiguous evidence that Joule heating transiently drives the magnetic medium above its Curie temperature, allowing magnetization reversal under a writing field of only \bw$\sim\pm$\SI{2}{\milli\tesla}, nearly 100 times smaller than the coercive field. The magnetic medium we employ in this study is Ni-intercalated TaSe$_{2}$, a two-dimensional (2D) itinerant metallic ferromagnet with \nitase composition. The magnetization direction in \nitase points along the out-of-plane direction (c-axis). This effective perpendicular magnetic anisotropy (PMA) is highly desirable for high-density magnetic recording. The use of intercalated 2D magnets presents an attractive pathway for low-temperature data storage technologies, offering a vast design space through diverse combinations of layered hosts and magnetic intercalants. Such materials not only expand the portfolio of candidate materials for next-generation high-density magnetic memories, but also open opportunities to explore novel strongly correlated phenomena at the intersection of magnetism and low-dimensional physics.
\section{Results}
\begin{figure*}[t]
    \centering
    \includegraphics[width=2\columnwidth]{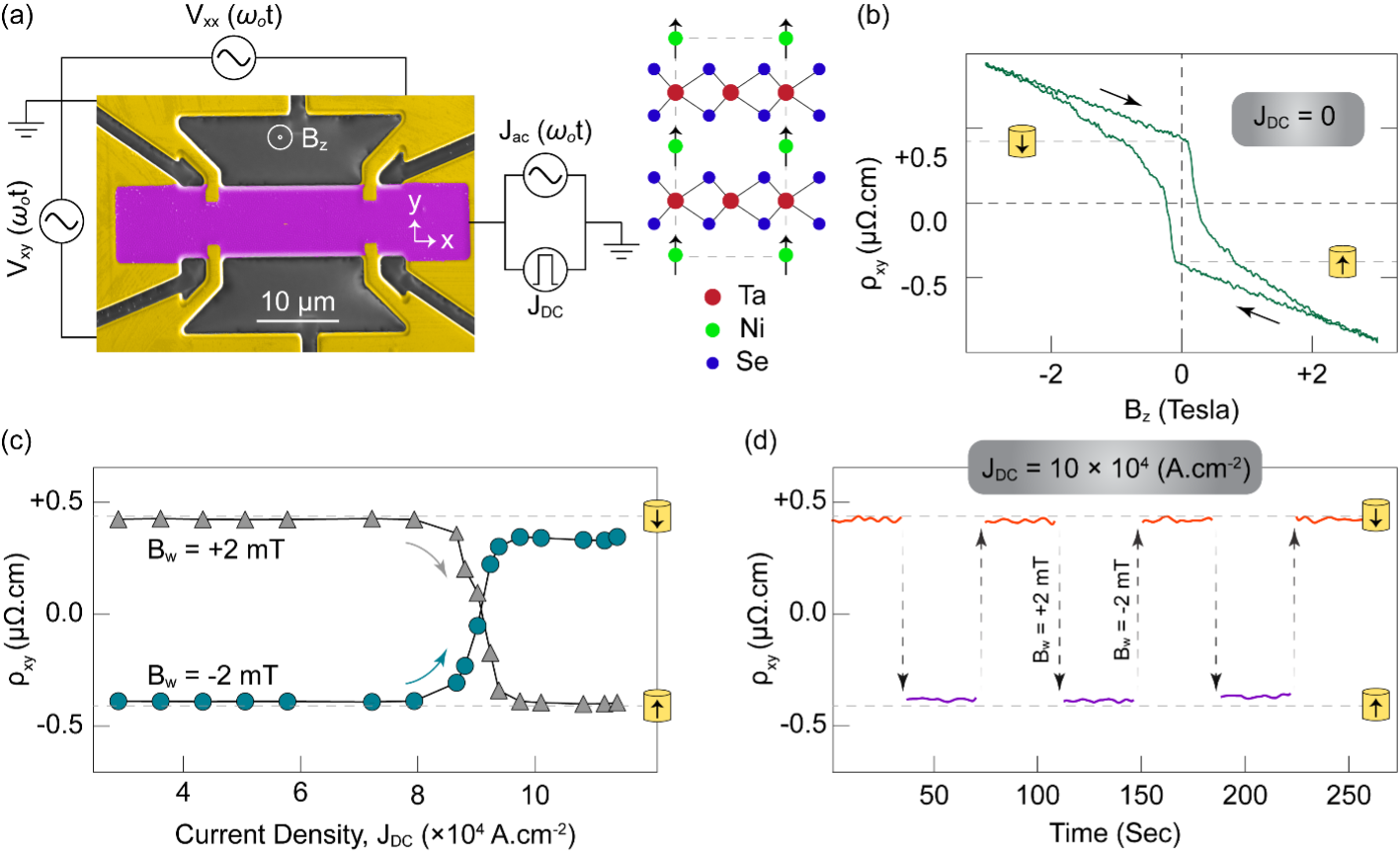}
    \caption{\textbf{Current-induced magnetization reversal in a Ni$_{1/4}$TaSe$_{2}$ FIB device.}
    \textbf{a} False colored SEM micrograph of a Ni$_{1/4}$TaSe$_{2}$ crystal patterned into a Hall bar. The magnetic field is applied along the out-of-plane direction (i.e., z-direction), which is the magnetic easy axis. Two current sources are used, a small ac component for electronic readout, and a pulsed DC current for magnetization reversal.
    \textbf{b} The anomalous Hall resistivity, showing distinct low- and high-resistance states, corresponding to the UP and DOWN magnetization states, respectively. 
    \textbf{c} Anomalous Hall resistivity measured after applying current pulses of various amplitudes. The device was initially set to a low (high) resistance state with a \SI{+3}{\tesla} (\SI{-3}{\tesla}) external field. A complete magnetization reversal is achieved at current densities above $10\times10^{4}$\si{\ampere}.\si{\per\square\centi\meter}. A writing field of \bw= $\pm$\SI{2}{\milli\tesla} was applied while pulsing.
    \textbf{d} Magnetization can be repeatedly reversed by only changing the sign of \bw under a fixed pulse current density of J$=10\times10^{4}$\si{\ampere}.\si{\per\square\centi\meter} $>$ J$_{\text{th}}$. Current pulses are \SI{1}{\milli\second} long. Measurements are done at \SI{2}{\kelvin} temperature. 
    }
    \label{fig:nitase}
\end{figure*}

We synthesized 2D \nitase crystals using a two-step chemical vapor transport method, as described in the Methods section. As schematically illustrated in Figure \ref{fig:nitase}(a), the material consists of layered basal planes of 2H-TaSe$_{2}$, with nickel intercalants occupying the Van der Waals gaps and aligning vertically with the tantalum atomic columns. As we have previously reported, \cite{Maksimovic2022_ItinerantMagnetism}, \nitase is a metallic ferromagnet with itinerant and local moment with a Curie temperature of T$_{\text{c}}\sim$ \SI{55}{\kelvin}, exhibiting a uniaxial magnetic anisotropy along the crystallographic c-axis, effectively mimicking PMA. For electronic transport measurements, high-quality crystals are patterned into Hall-bar devices using focused-ion beam (FIB) milling. A representative false-color scanning electron microscope image of a device is shown in Figure \ref{fig:nitase}(a), where the yellow regions correspond to gold contacts and the purple highlights the longitudinal transport channel. As schematically shown, the magnetization state of the device is probed by measuring the transverse resistivity (i.e., anomalous Hall effect, $\rho_{\text{xy}}$) and longitudinal resistivity (i.e., $\rho_{\text{xx}}$) in response to a small ac current applied along the longitudinal direction (x-direction). In current-assisted magnetization reversal experiments, a second current source delivers pulses of varying amplitude, while a writing field, \bw, is applied along the z-axis. Unless otherwise noted, all measurements were carried out at \SI{2}{\kelvin}.

First, we perform transport measurements under an external magnetic field to establish a reference for interpreting the current-assisted magnetization reversal experiments. Figure 2(b) shows the transverse resistivity of the device, which contains contributions from the anomalous Hall effect (i.e., the hysteretic component arising from the static magnetism) and the ordinary Hall effect (i.e., the linear-in-field background) over the field range of \SI{-3}{\tesla} to \SI{+3}{\tesla}. We distinguish two distinct resistivity states at \rhoxy$\sim\pm0.3$\si{\micro\ohm}.\si{\centi\meter}, corresponding to the DOWN and UP saturated magnetic states of Ni$_{1/4}$TaSe$_{2}$. From these measurements, we extract a coercive field of $\sim$\SI{0.2}{\tesla} where the AHE switches sign, while $\sim$\SI{2}{\tesla}, is required for complete magnetization reversal. The hysteresis loop is broad, reflecting the complex interplay of magnetic anisotropy and defects that pin domains.

To demonstrate the current-assisted magnetization reversal, we first initialize the device in the UP state using an external magnetic field of \SI{+3}{\tesla}. Then, we simultaneously apply a writing field of \bw = \SI{-2}{\milli\tesla} and \SI{1}{\milli\second} current pulses, J$_{\text{DC}}$, of varying amplitudes. After each current pulse, we record the \rhoxy  value. As shown in Figure \ref{fig:nitase}(c), a sharp transition occurs at a threshold current density of J$_{\text{th}}=10\times10^{4}$\si{\ampere}.\si{\per\centi\meter\squared}, where \rhoxy abruptly switches from $-0.3$\si{\micro\ohm}.\si{\centi\meter} to $+0.3$\si{\micro\ohm}.\si{\centi\meter}, indicating magnetization reversal. The final resistivity value above J$_{\text{th}}$ matches that of a fully saturated DOWN state obtained under an external magnetic field in Figure \ref{fig:nitase}(b), confirming complete UP-to-DOWN reversal. An identical threshold current density is observed when the device is initialized in the DOWN state with \SI{-3}{\tesla} and pulsed under \bw = \SI{+2}{\milli\tesla}, resulting in full reversal to the UP state (triangular symbols in Figure \ref{fig:nitase}(c)). This establishes that reversible, back-and-forth magnetization switching can be reliably achieved by simply reversing the polarity of the writing field while maintaining a fixed pulse amplitude above J$_{\text{th}}$. Such reversible switching is demonstrated in Figure \ref{fig:nitase}(d). Notably, the writing field of \SI{\pm2}{\milli\tesla} is two orders-of-magnitude smaller than the measured coercive field of \SI{0.2}{\tesla} (and 1000 smaller than the field required for complete sign reversal), clearly highlighting that the reversal is primarily assisted by the applied current pulses.

\begin{figure*}[t]
    \centering
    \includegraphics[width=2\columnwidth]{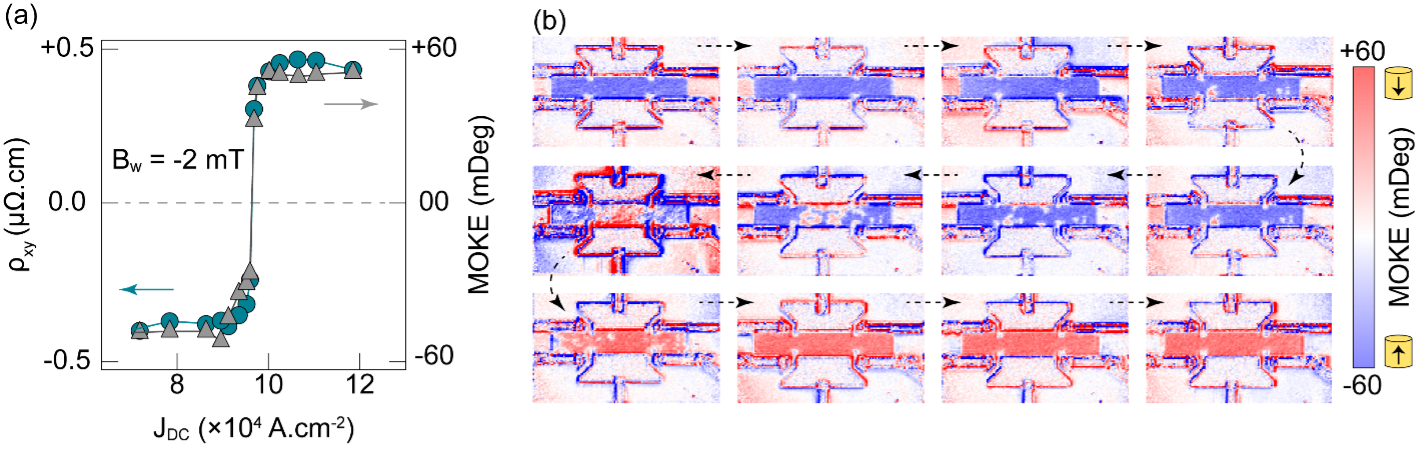}
    \caption{\textbf{Direct imaging of current-induced magnetization reversal using MOKE.}
    \textbf{a} Current-induced reversal probed electrically via anomalous Hall resistivity ($\rho_{\text{xy}}$, left axis) and optically via Kerr rotation (MOKE, right axis). Both readout modalities consistently yield a threshold current density of J$_{\text{th}} \sim 9.7\times10^{4}$\si{\ampere}.\si{\per\square\centi\meter}.
    \textbf{b} Sequence of Kerr microscopy maps acquired after successive current pulses of increasing amplitudes, corresponding to the data points presented in panel (a). The device is initialized to the UP state under a +\SI{+3}{\tesla} magnetic field and then pulsed under a writing field of B$_{w}$ = \SI{-2}{\milli\tesla}. Above the critical current density (J $>$ J$_{\text{th}}$), the MOKE contrast switches sign, directly confirming magnetization reversal from UP to DOWN.
    }
    \label{fig:moke}
\end{figure*}

We further provide direct evidence of current-induced magnetization reversal using magneto-optical Kerr effect (MOKE) microscopy, which enables the real-space visualization of the magnetic state of the \nitase under current pulses. As described in the Methods section, the polarization rotation of linearly polarized light upon reflection from the crystal (i.e., Kerr rotation) directly tracks the magnetization direction. Thus, UP and DOWN magnetization directions produce polarization rotations of opposite signs, allowing MOKE measurements to spatially map the magnetization state across the device. Following the protocol used in Figure \ref{fig:nitase}(c), we initialize the device in the UP state using an external field of \SI{+3}{\tesla}, then we apply a writing field of \bw = \SI{-2}{\milli\tesla} while sending current pulses of varying amplitudes. After each pulse, we simultaneously record the Kerr rotation and the anomalous Hall resistivity, and we plot them together in Figure \ref{fig:moke}(a). One can immediately see that the jump in the Hall resistivity at J$_{\text{th}}$ is accompanied by the sharp change in the sign of the Kerr rotation, confirming the magnetization reversal as the origin of the step-like response in both optical and electronic readouts. 

Leveraging the Kerr contrast, we spatially map the magnetic configuration of \nitase across the Hall bar while gradually increasing the current density from below to above J$_{\text{th}}$ (Figure \ref{fig:moke}(b)). Above the threshold current, the device switches into a spatially uniform DOWN state across the entire channel, confirming a complete magnetization reversal. Once saturation is reached, additional current pulses induce no further change (bottom row, Figure \ref{fig:moke}(b)). Importantly, the Kerr maps around J$_{\text{th}}$ reveal the underlying reversal mechanism: magnetization switching begins with the nucleation of small domains of opposite polarity, followed by their growth and coalescence into a fully saturated monodomain state. Such a nucleation – growth mechanism is the characteristic of magnets with macroscopic dimensions, consistent with our Hall bar geometry of $1.8 \times 9.5 \times 47 \si{\micro\meter\cubed}$.

\begin{figure}
    \centering
\includegraphics[width=1.0\columnwidth]{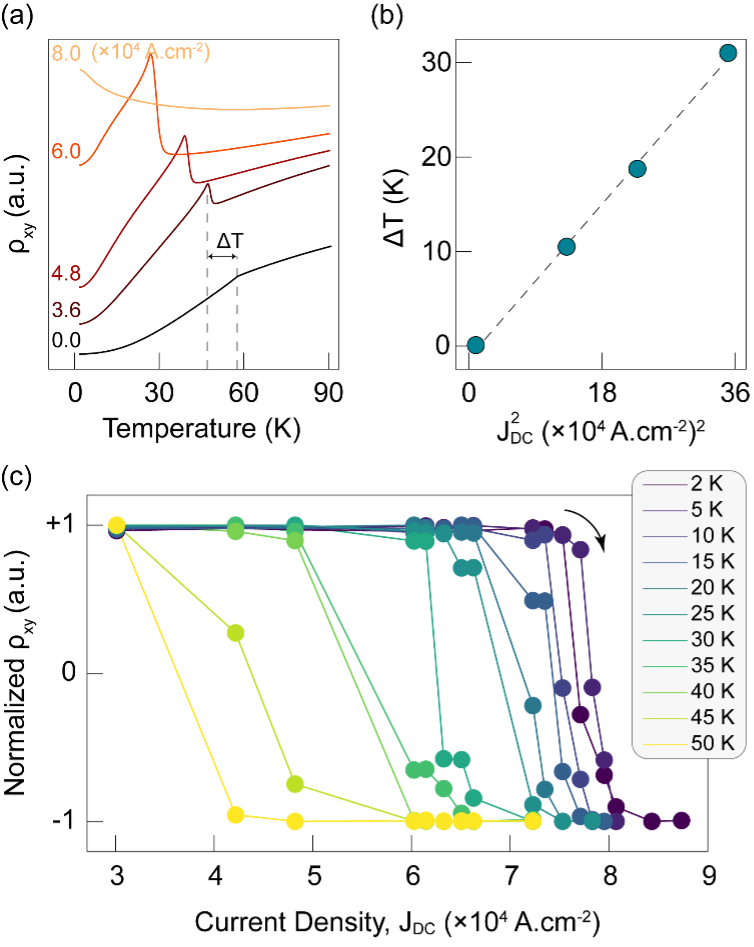}
    \caption{\textbf{Direct measurement of Joule heating.}
    \textbf{a} The temperature dependence of the longitudinal resistance \rhoxx in response to a small measuring ac current of J$_{\text{AC}}=20$\si{\ampere}.\si{\per\square\centi\meter}$\times \cos(2\pi f_{\circ}t)$ superimposed on a large DC current (J$_{\text{DC}}$) of varying amplitudes as indicated on the figure. The total applied current is J$_{\text{total}}$ = J$_{\text{AC}}$ + J$_{\text{DC}}$, with \rhoxx calculated from the measurement of the ac component of V$_{\text{xx}}$ at $f_{\circ}$ frequency.  \textbf{b} The shift of the apparent Curie temperature, T$_{c}$ (J$_{\text{DC}}$ $\neq$ 0), relative to the equilibrium Curie temperature, T$_{c}$ (J$_{\text{DC}}$ = 0), is defined as $\Delta$T = T$_{\text{c}}$ (J$_{\text{DC}}\neq$0) - T$_{\text{c}}$ (J$_{\text{DC}}$ = 0). $\Delta$T quadratically scales with J$_{\text{DC}}$, which is the signature of Joule heating. The absolute value of $\Delta$T is plotted. \textbf{c} Scaling of the switching-current density with temperature. A smaller J$_{\text{th}}$ is required at higher temperatures. 
    } 
    \label{fig:RIdc}
\end{figure}

Having established the current-assisted magnetization reversal and its switching mode, we next examine the role of assisting current in the switching mechanism. We first exclude torque-based contributions, \cite{Miron2011_SOT_Switching} primarily because (i) no asymmetric shift appears in the $\rho_{\text{xy}}$–B$_{\text{z}}$ hysteresis loop in the presence of a DC switching current, a signature of torque-based switching (see Figure \ref{fig:RvH_Idc}) \cite{Wu2019_SOT_fromFM,Fukami2016_SOT_AFMMFM}, and (ii) reversing the current polarity does not alter the switching characteristics (see Figure \ref{fig:Polarity}). We believe that the Joule-heating due to the current pulses during the write cycle is the main mechanism at play. To unambiguously probe this effect, we measure the temperature-dependence of the longitudinal resistivity, $\rho_{\text{xx}}$, in the presence of a DC current bias. As shown in Figure \ref{fig:RIdc}(a), we drive the device with a total current density

\begin{equation}
J_{\text{total}} = J_{\text{DC}} + J_{\text{AC}} \cos(2\pi f_{\circ}t)
\label{eq:Jtot}
\end{equation}
where the probe current is much less than the applied constant DC current, J$_{\text{AC}}$ $\ll$ J$_{\text{DC}}$.

Using lock-in detection technique at the ac-drive frequency of $f_{\circ}=$ \SI{37.77}{\hertz}, we measure the V$_{\text{xx}}$, from which we calculate $\rho_{\text{xx}}$. At J$_{\text{DC}}$ = 0, the temperature-dependence of $\rho_{\text{xx}}$ exhibits a kink at \SI{55}{\kelvin}, corresponding to the equilibrium Curie temperature of \nitase compound. With increasing J$_{\text{DC}}$, the kink shifts progressively to lower values, which we refer to as the “apparent” Curie temperature. Such a downshift indicates that the \nitase must be cooled below its equilibrium Curie temperature of \SI{55}{\kelvin} for an ordered state to be established in the crystal. At current density approaching $10\times10^{4}$\si{\ampere}.\si{\per\square\centi\meter}, the apparent Curie temperature approaches zero, implying that long-range ferromagnetic order cannot be sustained at any accessible temperature. Importantly, this critical current density of $8\times10^{4}$\si{\ampere}.\si{\per\square\centi\meter} coincides with the threshold current density J$_{th}\sim 9\times10^{4}$\si{\ampere}.\si{\per\square\centi\meter} identified independently from transverse resistivity $\rho_{\text{xy}}$ and MOKE measurements shown in Figure \ref{fig:nitase} and figure \ref{fig:moke}. These results provide unambiguous evidence that, at J$_{\text{th}}$, the current pulses transiently heat the \nitase medium above its Curie temperature. During this short-lived paramagnetic window, a small writing field of \bw = \SI{\pm2}{\milli\tesla}, 100 times smaller than the coercive field, is sufficient to reverse the magnetization. This experiment, therefore, constitutes direct proof of Joule-heating-assisted magnetization reversal.

As shown in Figure \ref{fig:RIdc} (b), the shift of the apparent Curie temperature relative to its equilibrium value given by,
\begin{equation}
\Delta T = T_{\text{c}} (J_{\text{DC}} \neq 0) - T_{\text{c}}(J_{\text{DC}} = 0)
\label{eq:delT}
\end{equation}
scales quadratically with the applied current density, that is $\Delta$T $\propto$ (J$_{\text{DC}}$)$^{2}$. This quadratic dependence stems from the Joule heating and explains the insensitivity to the polarity of the current pulses (Figure \ref{fig:Polarity}). In the context of EAMR, the extra energy supplied by the current pulses follows

\begin{equation}
E \propto\tau_{\text{pulse}}\times R\times(J_{\text{DC}})^{2}
\label{eq:EAMR}
\end{equation}
where R and $\tau_{\text{pulse}}$ are the device resistance and the pulse width, respectively. At base temperature of T$_{\circ}$ = \SI{2}{\kelvin}, a current density of J$_{\text{th}}$ $\sim9\times10^{4}$\si{\ampere}.\si{\per\square\centi\meter} increases the temperature of \nitase device by $\sim$ \SI{53}{\kelvin} (i.e., \SI{55}{\kelvin} – T$_{\circ}$ $\sim$ \SI{53}{\kelvin}). Accordingly, at higher measurement temperatures, smaller temperature increases are required to reach the Curie point, and thus lower current densities are sufficient for complete magnetization reversal. This technologically favorable scaling is experimentally demonstrated in Figure \ref{fig:RIdc}(c). Furthermore, the observed current-density dependence is quantitatively consistent with the equation \ref{eq:EAMR} particularly when accounting for the temperature dependence of the resistivity. We note that the DC current used in measurements of Figure \ref{fig:RIdc} were continuously applied (not pulsed), yet the extracted threshold current closely matches the value obtained from pulsed experiments, as discussed above. This agreement indicates that during the \SI{1}{\milli\second} current pulse the system reaches the thermal equilibrium. However, reducing the pulse width eventually increases the required threshold current density for magnetization reversal (Figure \ref{fig:PW_dep}).

 We note that the switching behavior reported in our work is different from existing reports on various intercalated 2D magnets where nuanced phenomena such as Rashba–Edelstein \cite {Nair2020} and exchange bias to glassy phase \cite {Maniv2021} are reported to generate two-level switching behaviors. Unlike previous reports, our work enables complete switching between two fully saturated magnetic states. Also notable is the current densities considered here, which are regarded low compared to related spintronic studies - orders of magnitude smaller than studies of spin-torque assisted switching, where routinely currents as large as $10^{8}$\si{\ampere}.\si{\per\square\centi\meter}  will be applied. \cite {Nguyen2024} Moreover, recent studies of related intercalated compounds Co$_{\text{x}}$TaS$_{2}$ \cite{Kim2025ElectricalControl3Q,Xiong2025all-electrical} and Ni$_{\text{x}}$NbS$_{2}$ \cite{Zhang2025HelicalSpinTexture} applied pulses this large over comparable timescales. Our study shows that currents orders of magnitude smaller can easily quench anneal the system above its transition temperature, allowing robust, reproducible responses, with tiny magnetic fields needed to create a deterministic switching of the AHE.

This study illustrates HAMR can be achieved entirely electrically utilizing Joule heating. The important characteristic of such systems is that they are required to have sufficiently large magnetic anisotropy to maintain thermally-stable non-volatile data storage, but can be softened by Joule heating to enable writing by very small magnetic fields. The \nitase system provides a balanced confluence of these requirements, demonstrating a mechanism whereby Joule heating at relatively small currents can be used to manipulate the magnetic polarization in fields that are orders of magnitude smaller than the coercive field. Our demonstration at low temperatures can be adopted by emerging cryogenic memory technologies, where robust and low-energy data storage is required for scaling both classical and quantum computing. \cite{Holmes2013Cryogenic,Baek2014HybridCryoMemory,Bardin2019CryoCMOSController}. However, the rapidly expanding material
space of intercalated 2D magnets leaves plenty of room for increasing ordering temperature by  judiciously combining 
various magnetic dopants and host materials.  Additionally, storing information directly in the spin degree of freedom naturally enables the seamless integration with spin-wave communication links, \cite{doi:10.1021/acs.nanolett.4c06592, Taghinejad2025, Chumak2015} paving the way toward full-stack spintronic platforms with unified logic and memory. 

\section{Methods}
\textbf{Crystal Growth.}High-quality single crystals of \nitase were grown by a two-step procedure. First, a precursor was prepared. The elements were combined in a ratio Ni:Ta:Se (x:1:2), where x = 0.25 with a 5$\%$ Se excess. The powders were loaded into an alumina crucible and sealed in a quartz tub under a partial pressure of argon gas (\SI{200}{Torr}). The tube was heated to \SI{900}{\celsius} and kept there for 5 days. The furnace was then shut off and allowed to cool naturally. This reaction yields a free-flowing shinny powder of crystallites. Second, \SI{5}{\gram} of such precursor was loaded with \SI{207}{\milli\gram} of iodine in a \SI{21}{\centi\meter} tube and placed in a horizontal two-zone furnace. The precursor and iodine were placed in one zone (zone1), while the growth occurred at the other end of the tube (zone2). Both zones were first heated to \SI{850}{\celsius} for 6 hours to encourage nucleation. Then, while zone 1 was maintained at \SI{850}{\celsius}, the temperature in zone 2 was reduced to \SI{750}{\celsius}. This condition was maintained for 5 days. The furnace was then shut off and allowed to cool naturally.

\textbf{Device Fabrication.} We used Ga$^{+}$ focused ion beam (FIB) to pattern the \nitase crystals into Hall-bar geometries for the transport measurements. Sputtered gold was deposited for electrical access to the device. Device was mounted on a measurement puck and wire bonded to the terminals of the puck for transport measurements.

\vspace{1mm}
\textbf{Transport Measurements.} A Keithley 6221 source meter was used to deliver current pulses along the longitudinal direction of the device. Longitudinal and transverse voltages were measured using lock-in detection in response to a \SI{37.77}{\hertz} AC current with an amplitude of \SI{50}{\micro\ampere}. The longitudinal ($\rho_{\text{xx}}$) and transverse ($\rho_{\text{xy}}$) resistivities were obtained by multiplying the measured voltages, V$_{\text{xx}}$ and V$_{\text{xy}}$, by the appropriate geometrical factor of the device and dividing by the applied current amplitude. For $\rho_{\text{xy}}$, a background contribution (the average of the maximum and minimum V$_{\text{xy}}$ values) was subtracted from the measured signal. Two device geometries were investigated: $1.8 \times 9 \times 22 \si{\micro\meter\cubed}$ and $1.5 \times 10 \times 20 \si{\micro\meter\cubed}$. All measurements were performed under vacuum in a Quantum Design PPMS cryostat, enabling precise control of temperature and magnetic field.

\vspace{1mm}
\textbf{MOKE Imaging.} MOKE imaging was performed in an optical cryostat (Quantum Design, OptiCool) with a 7T superconducting magnet and a base temperature of 2 K. Considering the PMA of 2D crystals, we adopted a polar MOKE geometry.The sample was illuminated with a red light-emitting diode (Thorlabs M617L3, center wavelength 617 nm), and the reflected light was collected by a 20× Mitutoyo objective and imaged onto a camera. The incident beam was linearly polarized prior to reaching the sample, and the reflected light passed through a second linear polarizer (analyzer). The analyzer was rotationally scanned near the crossed-polarization condition to extract the Kerr rotation angle. To eliminate any residue Kerr rotation angle from the imaging system, a MOKE image was taken on a large gold pad near the device and was used for background subtraction.

\vspace{3mm}
\section{Data Availability}
\textbf{Data Availability.} The datasets generated and/or analyzed during the current study are available from the corresponding author on reasonable request.

\vspace{3mm}
\section{Competing interests}
\textbf{Competing Interests.} All authors declare no financial or non-financial competing interests.

\vspace{3mm}
\section{Author Contributions}
JGA, HT conceived of the experiment. FW and RQ performed the MOKE measurements. JR and HT executed all transport experiments. JR made all the devices using FIB. JR and CX synthesized all the materials. All authors contributed to writing the manuscript.

\vspace{3mm}
\section{Acknowledgments}
\textbf{Acknowledgment.} This work was supported by the U.S. Department of Energy (DOE), Office of Science, Basic Energy Sciences, Materials Sciences Division, under Contract DEAC0205-CH11231 within the Quantum Materials program (KC2202). J.G.A. and H.T. acknowledge financial support from the Bakar Institute through the Bakar Prize. H.T. acknowledges the financial support of the Kavli Energy NanoScience Institute (ENSI) through the Heising-Simons Postdoctoral Fellowship at the University of California, Berkeley. J.G.A. acknowledges support of the HeisingSimons Faculty award H5881. 

\newpage

\bibliographystyle{unsrt}
\bibliography{magnetic_switching_refs}
\clearpage

\widetext
\begin{center}
\textbf{\large Supplemental Materials: Joule-Heating Assisted Magnetization Reversal in Intercalated 2D Magnets}
\end{center}
\setcounter{equation}{0}
\setcounter{figure}{0}
\setcounter{table}{0}
\setcounter{page}{1}
\makeatletter
\renewcommand{\theequation}{S\arabic{equation}}
\renewcommand{\thefigure}{S\arabic{figure}}
\renewcommand{\bibnumfmt}[1]{[#1]}
\renewcommand{\citenumfont}[1]{SM#1}
\maketitle
\section{Hall effect with applied dc current }
To rule out the role of spin–orbit torque (SOT)–induced switching mechanisms, we measured the Hall resistivity ($\rho_{\text{xy}}$) under an applied DC bias current superimposed on the low-frequency AC probe signal (\SI{70}{\micro\ampere}). The DC current was incrementally varied in both polarities while recording the anomalous Hall hysteresis loops as a function of the out-of-plane magnetic field. This approach allows us to examine whether the Hall loop center exhibits a current-dependent shift, which would indicate an effective SOT-induced field acting on the magnetic order. The presence of such a loop shift, or equivalently a bias-dependent change in coercivity, serves as a hallmark of current-driven spin–orbit torque effects, distinguishing them from purely thermal or magnetoresistive responses.
\begin{figure}[!htbp]
    \centering
    \includegraphics[width=0.85\columnwidth]{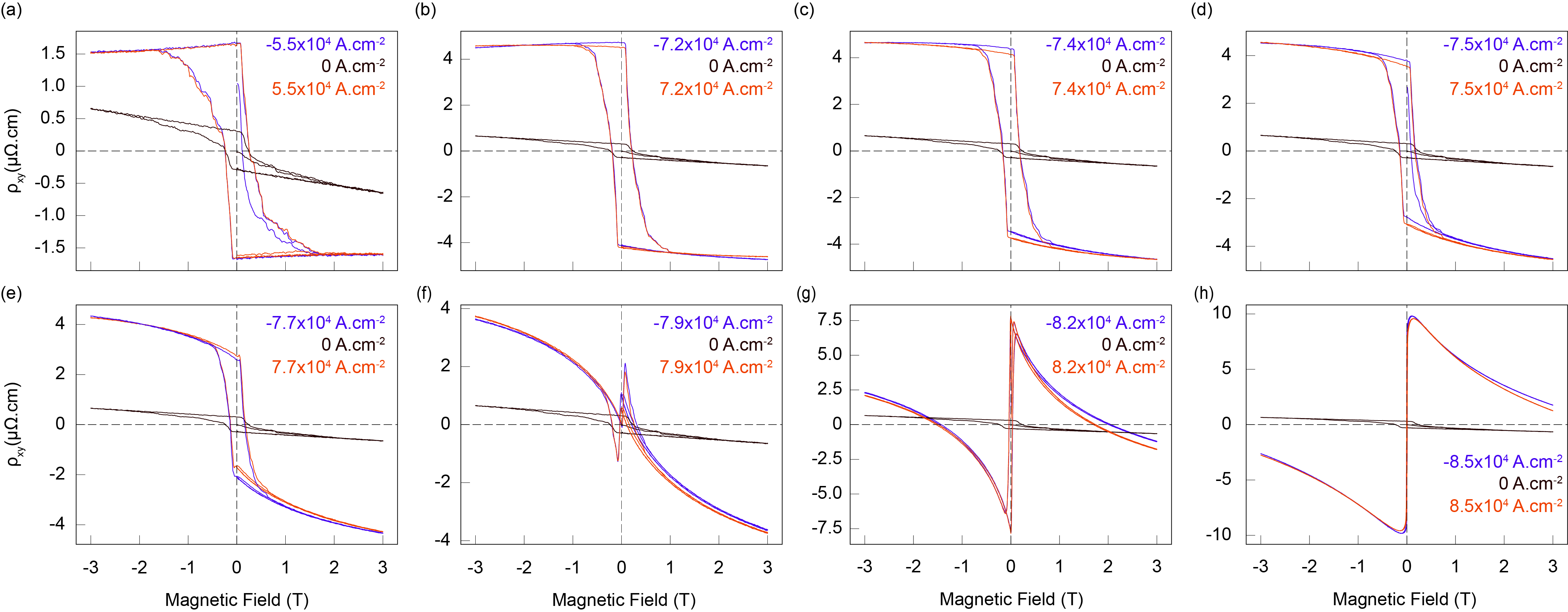}
    \caption{\textbf{Measured Hall effect under a perpendicular magnetic field with an applied positive and negative DC current.} Anomalous Hall resistivity measured (at \SI{2}{\kelvin}) with various applied DC currents. 
    }
    \label{fig:RvH_Idc}
\end{figure}

Figure \ref{fig:RvH_Idc} shows the measured Hall effect with various applied DC currents. There is no current-dependent shift in the hysteresis loop at any values of the applied DC current. Ruling out the possibility of SOT-induced switching mechanism. 


\section{Pulse current polarity}
\begin{figure}[!htbp]
    \centering
    \includegraphics[width=0.5\columnwidth]{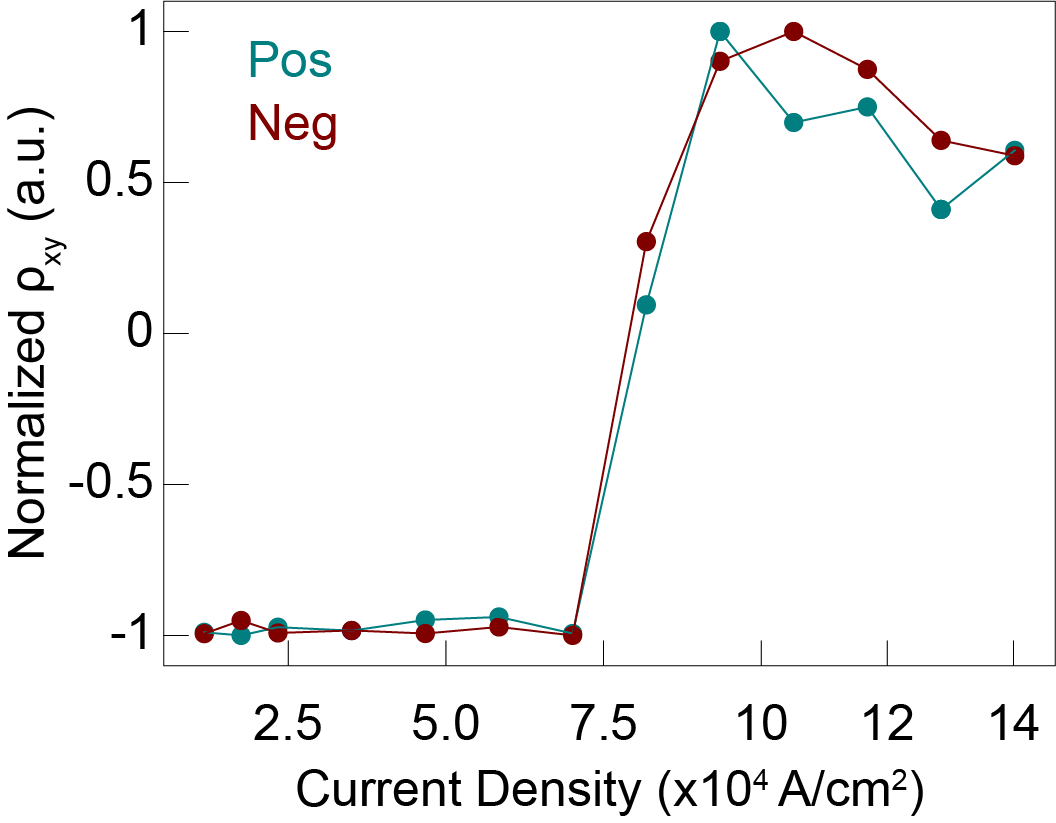}
    \caption{\textbf{Pulse current polarity dependence on the switching response.} Normalized anomalous Hall resistivity measured (at \SI{2}{\kelvin}) after applying current pulses of opposite polarity. The device was initially set to a low resistance state with a \SI{+3}{\tesla} external field. Reversing the polarity has no effect on the switching response.
    }
    \label{fig:Polarity}
\end{figure}

We start by preparing  the device in the low-resistance state by the application of a \SI{+3}{\tesla} external field. Then we apply current pulses of various amplitudes with a positive polarity (cyan curve) to achieve a switching event. Then the system was re-prepared in the low-resistance state by the application of \SI{+3}{\tesla} external field, followed by application of current pulses of negative polarity (burgundy curve). Reversing the polarity of the current pulses had no effect on the switching response (Figure \ref{fig:Polarity}). 

\section{Pulse Width dependence}
\begin{figure}[!htbp]
    \centering
    \includegraphics[width=0.5\columnwidth]{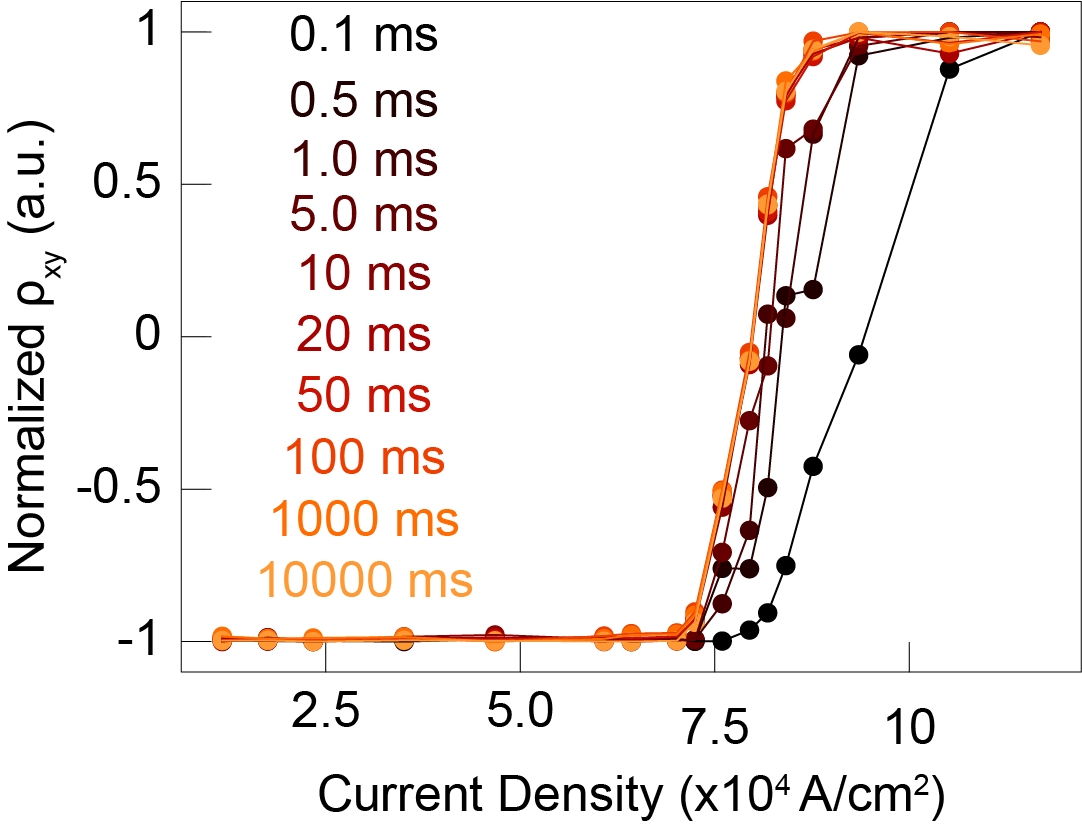}
    \caption{\textbf{Pulse width dependence on the switching response.} Normalized anomalous Hall resistivity measured (at \SI{2}{\kelvin}) after applying current pulses of various amplitudes and various pulse durations. The device was initially set to a low resistance state with a \SI{+3}{\tesla} external field. Reducing the pulse duration eventually increases the required threshold current density for magnetization reversal.
    }
    \label{fig:PW_dep}
\end{figure}

Figure \ref{fig:PW_dep} shows the dependence of the pulse width on the switching response. The device was first initialized to a low-resistance state with a \SI{+3}{\tesla} external field. Pulse currents of various amplitudes and pulse widths were applied to the device. As the pulse width approaches \SI{0.1}{\milli\second} the required threshold pulse current increases (black curve). At much higher pulse widths the threshold pulse current remains constant.

\end{document}